\documentclass[sigplan,9pt,nosumlimits]{acmart}
\makeatletter
\renewcommand\@formatdoi[1]{\ignorespaces}
\makeatother

\usepackage{booktabs} 
\usepackage{xcolor}

\usepackage{listings}
\usepackage[export]{adjustbox}
\usepackage{mdframed}
\usepackage{tcolorbox}
\usepackage{collectbox}
\usepackage{fancybox}
\usepackage{tabularx} 
\usepackage{todonotes}
\usepackage[linesnumbered, ruled]{algorithm2e} 
\SetAlFnt{\sffamily}
\usepackage{balance}
\usepackage{subcaption}

\usepackage{bbding}
\usepackage{pifont}
\usepackage{wasysym}
\usepackage{amssymb}
\usepackage{amsmath}
\usepackage{float}

\usepackage{tikz}
\usetikzlibrary{tikzmark}
\usetikzlibrary{positioning}
\usetikzlibrary{shapes.callouts}

\tikzset{
  level/.style   = { ultra thick, blue },
  connect/.style = { dashed, red },
  notice/.style  = { draw, rectangle callout, callout relative pointer={#1} },
  label/.style   = { text width=2cm }
}

\mdfdefinestyle{MyFrame}{%
    linecolor=black,
    outerlinewidth=0pt,
    linewidth=0.5pt,
    roundcorner=20pt,
    innertopmargin=02pt,
    innerbottommargin=02pt,
    innerrightmargin=02pt,
    innerleftmargin=02pt,
    backgroundcolor=white}

\usepackage[font=small,labelfont=bf]{caption}
\usepackage{varwidth}
\usepackage{algpseudocode}

\definecolor{ttblue}{rgb}{0,0,0.75}

\setcopyright{rightsretained}

\setcopyright{rightsretained}
\acmPrice{}
\acmDOI{}
\acmYear{}
\acmISBN{}
\acmConference[ACM SRC Grand Finals]{}{2019}{San Francisco, CA, USA}
\setcopyright{none}

\begin{document}
\title[\textsc{DynaSOAr}]{SPLASH: G: \textsc{DynaSOAr}: Accelerating Single-Method Multiple-Objects Applications on GPUs}
\titlenote{SPLASH 2018 ACM Student Research Competition, Graduate Category. Changed project name from \emph{SoaAlloc} to \emph{DynaSOAr}.}

\author{Matthias Springer}
\authornote{Academic Advisor: Hidehiko Masuhara, Tokyo Institute of Technology, Japan, \href{mailto:masuhara@acm.org}{masuhara@acm.org}}
\affiliation{%
  \institution{Tokyo Institute of Technology, Japan}
}
\email{matthias.springer@acm.org}


\begin{abstract}
Object-oriented programming (OOP) has long been regarded as too inefficient for SIMD high-performance computing, despite the fact that many important HPC applications have an inherent object structure. We discovered a broad subset of OOP that can be implemented efficiently on massively parallel SIMD accelerators. We call it \emph{Single-Method Multiple-Objects} (SMMO), because parallelism is expressed by running a method on all objects of a type.

To make fast GPU programming available to domain experts who are less experienced in GPU programming, we developed \textsc{DynaSOAr}, a CUDA framework for SMMO applications. \textsc{DynaSOAr} improves the usage of allocated memory with an SOA  data layout and achieves low memory fragmentation through efficient management of free and allocated memory blocks with lock-free, hierarchical bitmaps.
\end{abstract}


%
%


\keywords{Hierarchical Bitmaps, GPUs, Memory Allocation, OOP}

\maketitle

\section{Introduction}
General-purpose GPU computing has long been a tedious job, requiring programmers to write hand-optimized, low-level programs. In an attempt to make GPU computing available to a broader range of developers, our efforts are centered around bringing fast object-oriented programming (OOP) to low-level languages such as CUDA and C++.

OOP has a wide range of applications in high-performance computing (HPC) but is often avoided due to bad performance~\cite{master_th_patel}. Dynamic memory management (DMM), i.e., the ability/flexibility of creating/deleting objects at any time, is one of the corner stones of OOP. Programmers often have to go to great lengths to build their own allocators or write their applications is a more convoluted way due to slow DMM (e.g.,~\cite{doi:10.1002/cpe.3808}).

In recent years, fast, dynamic memory allocators have been developed for GPUs~\cite{6339604, hallocweb, Gelado:2019:TGM:3293883.3295727}. However, while these allocators often provide good raw (de)allocation performance, they miss key optimizations for structured data, leading to poor data locality and memory bandwidth utilization when accessing allocated memory.

\subsection{Single-Method Multiple-Objects (SMMO)}
We identified a class of HPC applications that can be expressed as object-oriented programs and implemented efficiently on SIMD architectures. We call this class \emph{Single-Method Multiple-Objects} (SMMO) because parallelism is expressed by running a method on all objects of a type (\emph{parallel do-all}). Objects may be created/deleted at any time within a parallel do-all operation. 


SMMO is a broad class of problems\footnote{We implemented a few SMMO applications from different domains: \url{https://github.com/prg-titech/dynasoar/wiki/Benchmark-Applications}} with many real-world applications~\cite{DBLP:journals/corr/abs-1810-11765}, such as simulations for population dynamics, (e.g., Sugarscape~\cite{RePEc:mtp:titles:0262550253}), evacuations~\cite{doi:10.1002/cpe.3808}, wildfire spreading~\cite{doi:10.1080/21580103.2016.1262793}, finite element methods or particle systems, to name just a few. SMMO can also express BFS graph traversals and  dynamic tree updates/ constructions such as in Barnes-Hut~\cite{BURTSCHER201175}.

\subsection{Object Allocation with Efficient Memory Access}
We developed \textsc{DynaSOAr}~\cite{DBLP:journals/corr/abs-1810-11765}, a CUDA framework for SMMO applications. \textsc{DynaSOAr} consists of three parts:

\begin{enumerate}
  \item A \textbf{dynamic memory allocator} based on concurrent, lock-free, hierarchical bitmaps. Our allocator stores objects in a Structure of Arrays (SOA) data layout.
  \item An efficient \textbf{parallel do-all operation}. Such operations spawn a CUDA kernel and control the assignment of objects to GPU threads.
  \item An embedded C++ \textbf{data layout DSL} for object-oriented programming with SOA layout, inspired by \textsc{Ikra-Cpp}~\cite{Springer:2018:ICD:3178433.3178439}. This DSL allows us to implement \textsc{DynaSOAr} entirely in C++/CUDA without a custom code generator/preprocessor.
\end{enumerate}

\begin{figure*}
\includegraphics[width=\textwidth]{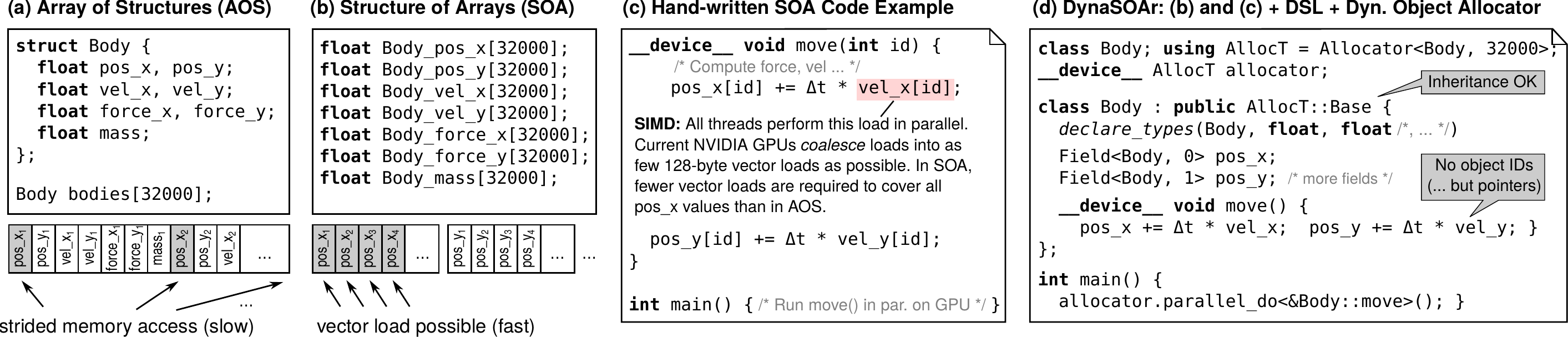}
\caption{N-Body example: Excerpt of structure definition and Euler method time step (AOS/SOA/\textsc{DynaSOAr}).}
\label{fig:nbody_aos_soa}
\end{figure*}

\textsc{DynaSOAr} achieves significant speedups over state-of-the-art GPU memory allocators by controlling both (1) data layout and (2) data access.

\paragraph{Structure of Arrays (SOA) Layout}
SOA is a well-studied best practice of SIMD programming. Compared to a traditional AOS layout (Fig.~\ref{fig:nbody_aos_soa}a), an SOA layout (Fig.~\ref{fig:nbody_aos_soa}b) can increase memory bandwidth and cache utilization when the same fields of multiple objects are simultaneously accessed~\cite{HOMANN2018325}.

SIMD architectures achieve parallelism by running the same processor instruction on a vector register. Even though recent GPUs appear to have thousands of \emph{CUDA cores}, the hardware actually only has a few hundred physical cores, each operating on 128-byte vector registers containing 32 \texttt{float}/\texttt{int} scalars\footnote{We are focusing on recent NVIDIA architectures, but other architectures are similar.}. Memory accesses of a physical core (i.e., 32 consecutive GPU threads; \emph{warp} in CUDA) that fall into an aligned 128-byte address window are serviced with efficient vector loads (\emph{memory coalescing}). Getting data into and out of vector registers is often the biggest bottleneck and peak memory bandwidth utilization can be achieved only with memory coalescing.

SOA can increase memory coalescing when accessing structured data because values of the same field are stored together. We improved the SOA layout in three ways:
\begin{itemize}
  \item We built an embedded \textbf{C++/CUDA DSL} such that programmers can write code like in Fig.~\ref{fig:nbody_aos_soa}d instead of Fig.~\ref{fig:nbody_aos_soa}c.
  \item We developed the first \textbf{GPU dynamic memory allocator} (C++ \texttt{new}/\texttt{delete} interface) with SOA performance characteristics. \textbf{Low fragmentation} is key: If data is fragmented, more vector accesses are needed for the same number of bytes, reducing the benefit of SOA.
  \item Contrary to other systems such as \emph{Columnar Objects}~\cite{Mattis:2015:COI:2814228.2814230}, we support class inheritance without wasting memory.
\end{itemize}

\subsection{Contributions}
Our work contributes to the state-of-the-art as follows:
\begin{itemize}
  \item The concept of and examples of \textbf{SMMO} applications.
  \item \textbf{SOA Improvements}: Embedded C++ DSL and subclassing.
  \item A \textbf{dynamic memory allocator} for GPUs with efficient memory access and low fragmentation.
  \item A concurrent, lock-free, hierarchical \textbf{bitmap data structure}, based on atomic operations and retry loops.
\end{itemize}

\section{The \textsc{DynaSOAr} System}
\begin{figure}
  \centering
  \includegraphics[width=0.5\columnwidth]{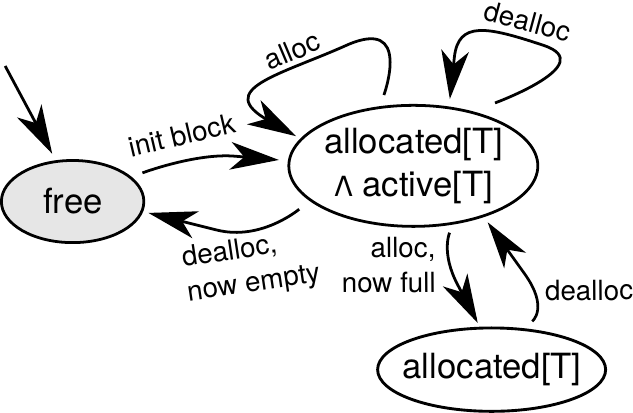}
  \caption{Block states. Initially, every block is \emph{free}. \emph{Allocated} blocks contain only objects of a specific type $T$. To reduce fragmentation, new objects are allocated in \emph{active} blocks of the corresponding type.}
  \label{fig:block_states}
\end{figure}
\begin{figure*}
  \centering
  \includegraphics[width=0.9\textwidth]{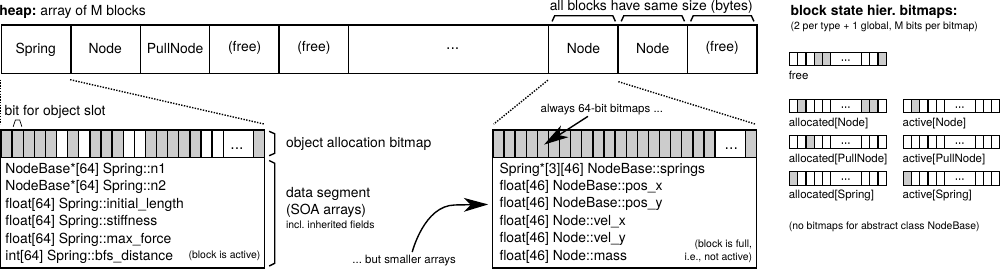}
  \caption{Heap layout of a finite element method. \textsc{DynaSOAr} is a \emph{slab allocator}~\cite{Bonwick:1994:SAO:1267257.1267263}: Only objects of the same C++ type are stored in a block.}
  \label{fig:heap_layout}
\end{figure*}

\textsc{DynaSOAr} divides the heap into blocks of equal size in bytes (Fig.~\ref{fig:heap_layout}). Every block contains only objects of the same C++ class/struct type in SOA layout: one \emph{SOA array} per field. Blocks of the smallest type in the system\footnote{All types must be pre-declared with the allocator (template arg. in 1st line of Fig.~\ref{fig:nbody_aos_soa}d).} always have a \emph{capacity} (\#object slots) of 64 and determine the block size in bytes, and thus the number of blocks $M$ on the heap. Blocks of other types may have a smaller capacity depending on the size of their type in relation to the smallest type.

\paragraph{Coalesced Access}
Although this layout does not constitute a plain structure of arrays as in Fig.~\ref{fig:nbody_aos_soa}b but \emph{multiple structures} of arrays (one per block), it has the same performance characteristics as long as every SOA array is at least 128~bytes big (cache line and vector register size). This is the case for block capacities of at least 32 (assuming 4~byte fields).

\paragraph{Block States}
A 64-bit \emph{object allocation bitmap} keeps track of allocations. A block can be in one or more \emph{multistates} (Fig.~\ref{fig:block_states}). 
\begin{itemize}
  \item \emph{free:} The block is empty and does not contain any objects.
  \item \emph{allocated[T]:} The block contains at least 1 object of type $T$. No other types can be stored in this block.
  \item \emph{active[T]:} The block contains objects of type $T$. It is not full yet, i.e., it has space for at least one more object. \\ \emph{active[T]} $\Rightarrow$ \emph{allocated[T]}.
\end{itemize}

Allocation and deallocation routines frequently have to lookup blocks by state. For that reason, block states are indexed by bitmaps of size $M$; one bitmap per state.

\paragraph{Concurrency}
Due to concurrent operations of other threads, these bitmaps can be temporarily inconsistent with the actual state of a block. Designing lock-free algorithms is notoriously difficult~\cite{Michael:2004:SLD:996841.996848} and we carefully designed our algorithms such that they can detect inconsistencies and retry if necessary.

Our allocator design has one key feature that makes this problem more tractable: Apart from the data segment, every block has the same structure. Object allocation bitmaps are always located at the same offset within a block and the memory locations of blocks are fixed. Therefore, our algorithms can use block state bitmaps to quickly lookup blocks but then update object allocation bitmaps, which are the \emph{single source of truth}, with atomic operations. At this \emph{linearization point}~\cite{Herlihy:1990:LCC:78969.78972}, our algorithms can detect inconsistencies\footnote{Due to space limitations, we cannot give an exhaustive description of our algorithms here, but we invite the interested reader to take a look at our full paper~\cite{DBLP:journals/corr/abs-1810-11765}.}.

\subsection{Bitmap Operations}
\textsc{DynaSOAr} uses large bitmaps for indexing block states. These bitmaps provide a few basic operations.
\begin{itemize}
  \item \textsf{try\_clear(pos)}: Atomically clear the bit at \textsf{pos}. If the bit was already cleared, return \textsf{false}, otherwise \textsf{true}.
  \item \textsf{clear(pos)}: Clear the bit at \textsf{pos}. This is identical to: \textsf{\textbf{while} (!try\_clear(pos)) \{\}}.
  \item \emph{Similarly:} \textsf{try\_set(pos)} and \textsf{set(pos)}.
  \item \textsf{try\_find\_set()}: Find and return the position of a set bit or \textsf{FAIL} if none was found (e.g., because the bitmap is inconsistent).
  \item \textsf{clear()}: Find a set bit and clear it. This is identical to: \textsf{\textbf{while} ((i = try\_find\_set()) != FAIL \&\& try\_clear(i)) \{\};} \textsf{\textbf{return} i;}
\end{itemize}

All operations are thread-safe and can be invoked by threads concurrently. \textsf{try\_clear(pos)} and \textsf{try\_set(pos)} are implemented with atomic bit-wise OR/AND operations\footnote{Atomic operations return the original value in memory.}. Atomic operations are generally slow, but became much faster with recent GPUs~\cite{DeGonzalo:2019:AGW:3314872.3314884}.

\subsection{Object Allocation}
Alg.~\ref{alg:alloc_algo} gives a simplified overview of object allocation. This is a thread-safe and entirely lock-free algorithm.

To keep memory fragmentation low, \textsc{DynaSOAr} allocates new objects always in active blocks. These are blocks that already contain some objects of the same type, but still have some vacancy. Only if no active block could be found, a new block is initialized (\emph{slow path}).

This technique is in contrast to state-of-the-art GPU allocators, which scatter new allocations in the heap (e.g., using hashing), to reduce collisions and synchronization among threads~\cite{6339604, hallocweb}. \textsc{DynaSOAr} requires more synchronization between threads, in the form of atomic operations. This makes (de)allocations more expensive. However, \textsc{DynaSOAr}'s allocation policy leads to much lower fragmentation and denser allocations. This results in better overall performance of SMMO applications on SIMD architectures.

\SetKwComment{Comment}{\textcolor{gray}{$\triangleright$\ }}{}
\SetAlgoVlined
\begin{algorithm}[t]
\small
 \Repeat(\Comment*[f]{\textcolor{gray}{\textsf{infinite loop if OOM}}}){false; \hfill \emph{\textsf{$\triangleright$\  \textcolor{gray}{select new block + retry}}}}{
  bid $\gets$ active[T].\emph{try\_find\_set}()\;
  \If(\Comment*[f]{\textcolor{gray}{\textsf{slow path}}}){\emph{bid} = FAIL} {
    bid $\gets$ free.\emph{clear}()\;
    \emph{initialize\_block}<T>(bid)\;
    allocated[T].\emph{set}(bid)\;
    active[T].\emph{set}(bid)\;
  }
  alloc $\gets$ \emph{get\_block}<T>(bid).\emph{reserve}()\;
  \If(\Comment*[f]{\textcolor{gray}{\textsf{block filled up after line~2?}}}){\emph{alloc} $\not=$ FAIL}{
    \If(\Comment*[f]{\textcolor{gray}{\textsf{allocated last obj. slot}}}){\emph{alloc.state} = FULL}{
      active[t].\emph{clear}(bid)\;
    }
    \Return \emph{make\_pointer}(bid, alloc.slot)\;
  }
 }
 \caption{Allocator::allocate<T>() : T* \hfill \fbox{GPU}}
 \label{alg:alloc_algo}
\end{algorithm}

\subsection{Parallel Do-all}
In \textsc{DynaSOAr}, we express parallelism of SMMO applications with parallel do-all operations. Such operations spawn a CUDA kernel and run a method in parallel for all objects of a type $T$. Objects are assigned to GPU threads in such a way that the GPU can coalesce field reads/writes of the \texttt{\textbf{this}} object\footnote{Field reads/writes of other objects do not benefit from additional memory coalescing.}. This increases memory bandwidth utilization and overall application performance.

\begin{figure}
  \centering
  \includegraphics[width=0.8\columnwidth]{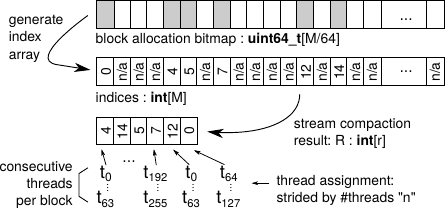}
  \caption{Compaction of \emph{allocated[T]} array and assigning objects to GPU threads. Assuming block capacity $n=64$.}
  \label{fig:thread_assignment}
\end{figure}

Before a CUDA kernel is launched, we precompute which objects are processed by each GPU thread (Fig.~\ref{fig:thread_assignment}) in two steps that can be implemented efficiently on GPUs.

\begin{enumerate}
  \item Generate an index array of set bits in \emph{allocated[T]}.
  \item Filter out unused index values (\emph{stream compaction}). Can be implemented with a prefix sum algorithm~\cite{IJNC151}.
\end{enumerate}

If $c$ is the capacity of blocks of type $T$, we assign $c$ consecutive threads to each allocated block of type $T$. Since these $c$ threads have consecutive IDs, they will run on the same 1--3 physical cores, so accesses are coalesced; not necessarily perfectly coalesced but better than in AOS. Not all blocks are entirely full, so some threads will have no objects to process; this mechanism is nevertheless fast, because we expect blocks to have a high average fill level since new allocations are always performed in active blocks.

Inside a CUDA kernel, a thread can now quickly read its assigned blocks from the filtered index array $R$ and process the objects at slot $\mathit{tid} \,\, \% \,\, c$ in these blocks, if they have an object allocated there.

\subsection{Additional Optimizations}
\textsc{DynaSOAr} employs optimizations in addition to SOA to speed up allocations.
\begin{itemize}
  \item Block bitmaps are \textbf{hierarchical} and updated with \textbf{eventually consistent, lock-free algorithms}, so that we can find \emph{free} blocks (\textsf{try\_find\_set()}) with a log. number of accesses.
  \item To reduce thread contention, we borrow \textbf{allocation request coalescing} from XMalloc~\cite{5577907}: A leader thread allocates objects on behalf of all threads in a warp.
  \item Efficient implementation and engineering efforts: We utilize \textbf{low-level optimizations} such as int. intrinsics (e.g., Alg~\ref{alg:clear}).
\end{itemize}

\section{Hierarchical Bitmaps}
\textsc{DynaSOAr} uses hierarchical bitmaps to find find free and active blocks with \textsf{try\_find\_set()} in Alg.~\ref{alg:alloc_algo}. This is a key optimization because block bitmaps can reach multiple megabytes in size and checking every bit is too slow.

\begin{figure}
  \centering
  \includegraphics[width=0.7\columnwidth]{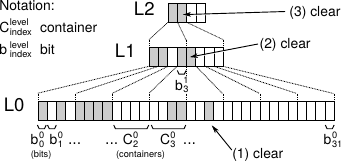}
    \caption{Example: Hier. bitmap of size $N=32$ with container size 4 (instead of 64). This example illustrates how (1) a \emph{clear}(18) operation triggers (2) a \emph{clear}(4) operation in the nested bitmap, which triggers (3) a \emph{clear}(1) operation in the next nested bitmap.} \label{fig:bitmap_clear_example}
\end{figure}

\paragraph{Data Structure}
A hierarchical bitmap of size $N$ bits consists of two parts: an array of size $\lceil N/64\rceil$ of 64-bit \emph{containers} (\texttt{uint64\_t}), and a \emph{nested bitmap} of size $\lceil N/64 \rceil$ if $N > 64$. A container $C_i^l$ consists of bits $b_{64 \cdot i}^l$, ...,  $b_{64 \cdot i + 63}^l$ and is represented by one bit $b_{i}^{l+1}$ in the nested (higher-level) bitmap (Fig.~\ref{fig:bitmap_clear_example}). That bit is set if at least one bit is set in the container.

\begin{align*}
b_i^{l+1} = \bigvee_{k=0}^{63} b_{64\cdot i + k}^{l} \tag{\emph{container consistency}}
\end{align*}

Bits in are changed with atomic operations. Higher-level bits (and thus bitmaps) are \emph{eventually consistent}\footnote{This is a key difference from other lock-free hier. data structures such as SNZI~\cite{Ellen:2007:SSN:1281100.1281106}, which have stronger runtime consistency guarantees and require complex algorithms.} with their containers. Keeping both consistent all the time is difficult without locking, because two different memory locations cannot be changed together atomically. However, due to the design of the bitmap operations, the bitmap is guaranteed to be in a consistent state when all bitmap operations (of all threads) are completed. Bitmap operations retry or give up (\emph{FAIL}) to handle temporary inconsistencies.

\paragraph{\textsf{clear(pos)} with Atomics}
Alg.~\ref{alg:clear} illustrates how to clear a specific bit. We clear the bit with an atomic operation (Line~4). If this operation actually changed the bit (\textsf{success}) and cleared the last bit of the container, it is this thread's \emph{responsibility} to clear the respective bit in the nested bitmap. Since the bitmap may be in an inconsistent state, we have to retry until the bit was cleared, thus \textsf{clear(pos)} instead of \textsf{try\_clear(pos)} in Line~7.

\SetKwComment{Comment}{$\triangleright$\ }{}
\begin{algorithm}[t]
\small
 cid $\gets$ pos / 64\;
 offset $\gets$ pos \% 64\;
 mask $\gets$ 1 {<}{<} offset\;
 prev $\gets$ \emph{atomicAnd}(\&container[cid], $\sim$mask)\;
 success $\gets$ (prev \& mask) $\not=$ 0\;

\If{\hspace{-4pt} success $\wedge$ has\_nested $\wedge$ {popc}(\emph{prev})\hspace{-0.17cm}\tikz[remember picture] \node [] (d){};\hspace{-2pt} \emph{\,\,=\,\,1}}{
  nested.\emph{clear}(cid)\;
}
\begin{tikzpicture}[remember picture, overlay,
    every edge/.append style = { ->, thick, >=stealth,
                                  dashed, line width = 1pt },
    every node/.append style = { align = center,
                                 font=\sffamily\small, fill= gray!20},
                  text width = 1.9cm ]
  \node [notice={(-0.2,0.15)}, below right = 0.3cm and -1.5cm of d]  (D) {\mbox{\begin{varwidth}{1.9cm} \scriptsize \textbf{population cnt.}: number of set bits\end{varwidth}}};

\end{tikzpicture} \hspace{-0.17cm} \textbf{return } success;
 \caption{Bitmap::\emph{try\_clear}(pos) : void  \hfill \fbox{GPU}}
 \label{alg:clear}
\end{algorithm}

\section{Evaluation}
\begin{figure*}
  \begin{subfigure}[t]{0.12\textwidth}
    \centering
    \includegraphics[width=\textwidth]{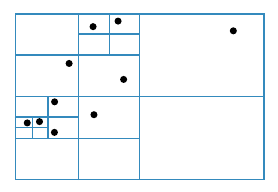}
    \vspace{-0.6cm}
    \caption*{\footnotesize \textbf{\textsf{(a)}} \textsf{barnes-hut}~\cite{BURTSCHER201175}: \\ Parallel Tree Constr.}  
  \end{subfigure}\hfill
  \begin{subfigure}[t]{0.12\textwidth}
    \centering
    \includegraphics[width=\textwidth]{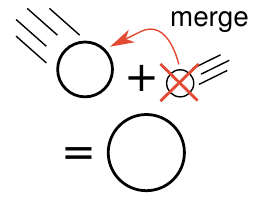}
    \vspace{-0.6cm}
    \caption*{\footnotesize \textbf{\textsf{(b)}} \textsf{collisions}: \\ Particle System} 
  \end{subfigure}\hfill
  \begin{subfigure}[t]{0.12\textwidth}
    \centering
    \includegraphics[width=\textwidth]{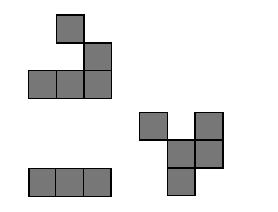}
    \vspace{-0.6cm}
    \caption*{\footnotesize \textbf{\textsf{(c)}} \textsf{game-of-life}: \\Cellular Automaton} 
  \end{subfigure}\hfill
  \begin{subfigure}[t]{0.12\textwidth}
    \centering
    \includegraphics[width=\textwidth]{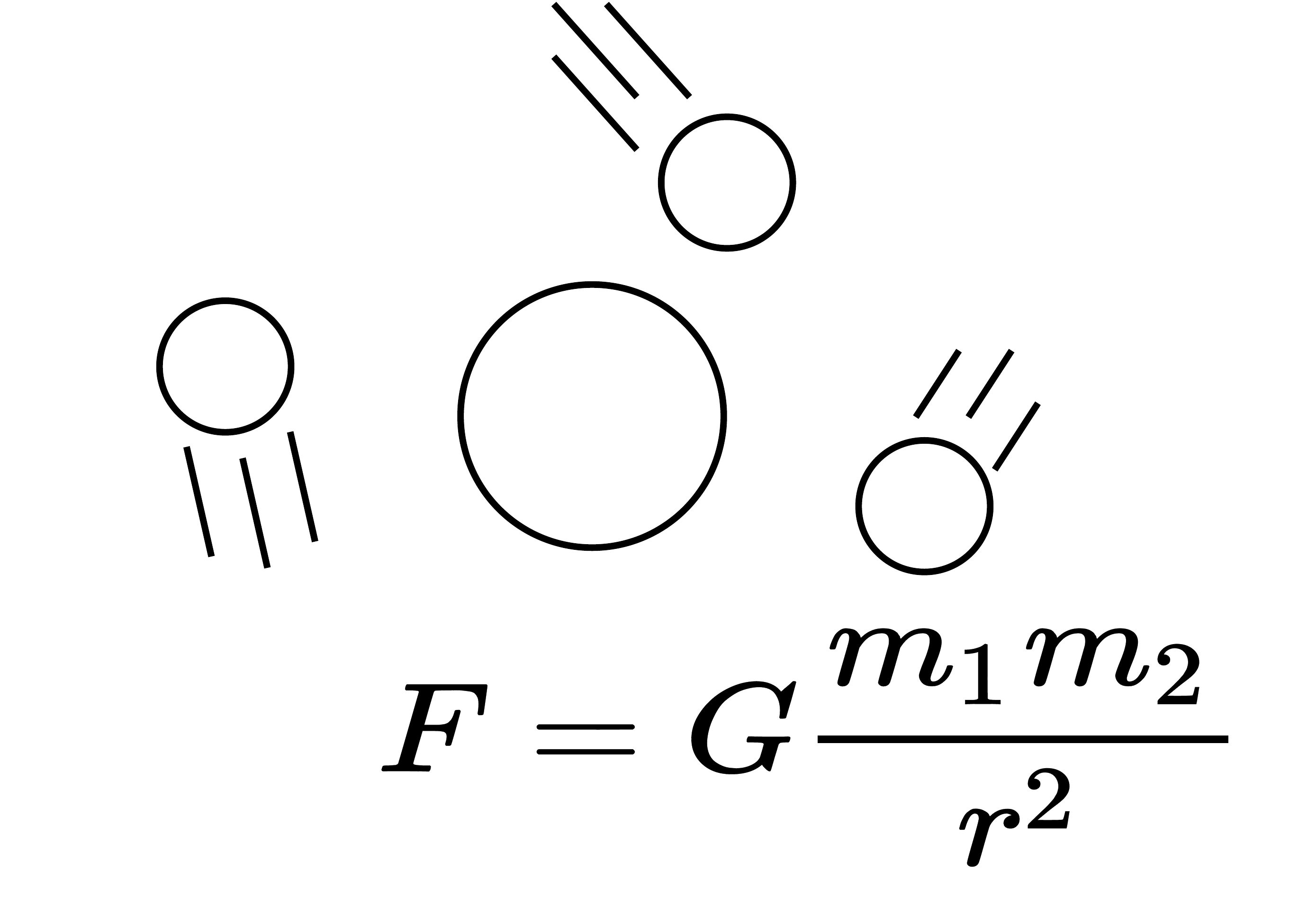}
    \vspace{-0.6cm}
    \caption*{\footnotesize \textbf{\textsf{(d)}} \textsf{nbody}: \\ Particle System} 
  \end{subfigure}\hfill
  \begin{subfigure}[t]{0.12\textwidth}
    \centering
    \includegraphics[width=\textwidth]{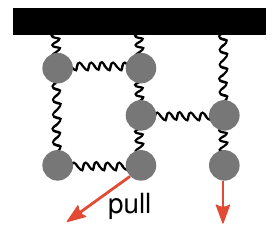}
    \vspace{-0.6cm}
    \caption*{\footnotesize \textbf{\textsf{(e)}} \textsf{structure}~\cite{LU2018240}: \\Finite Elem. Method}  
  \end{subfigure}\hfill
  \begin{subfigure}[t]{0.12\textwidth}
    \centering
    \includegraphics[width=\textwidth]{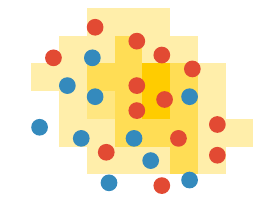}
    \vspace{-0.6cm}
    \caption*{\footnotesize \textbf{\textsf{(f)}} \textsf{sugarscape}~\cite{RePEc:mtp:titles:0262550253}:\\ Agent-based Sim.} 
  \end{subfigure}\hfill
  \begin{subfigure}[t]{0.12\textwidth}
    \centering
    \includegraphics[width=\textwidth]{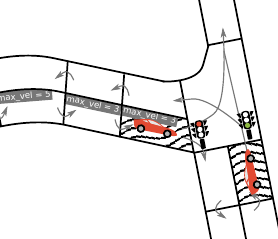}
    \vspace{-0.6cm}
    \caption*{\footnotesize \textbf{\textsf{(g)}} \textsf{traffic}~\cite{nagel_schr}: \\ Nagel-Schr. Model} 
  \end{subfigure}\hfill
  \begin{subfigure}[t]{0.12\textwidth}
    \centering
    \includegraphics[width=\textwidth]{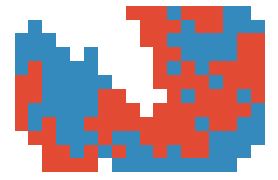}
    \vspace{-0.6cm}
    \caption*{\footnotesize \textbf{\textsf{(h)}} \textsf{wator}~\cite{10.2307/24969495}: \\ Agent-based Sim.} 
  \end{subfigure}
  \caption{Visualization and classification of SMMO benchmark applications. The SMMO structure of each app. is explained in detail in the GitHub wiki page\protect\footnotemark[1].}
  \label{fig:smmo_apps}
\end{figure*}

\begin{figure*}
  \centering
  \includegraphics[width=0.85\textwidth]{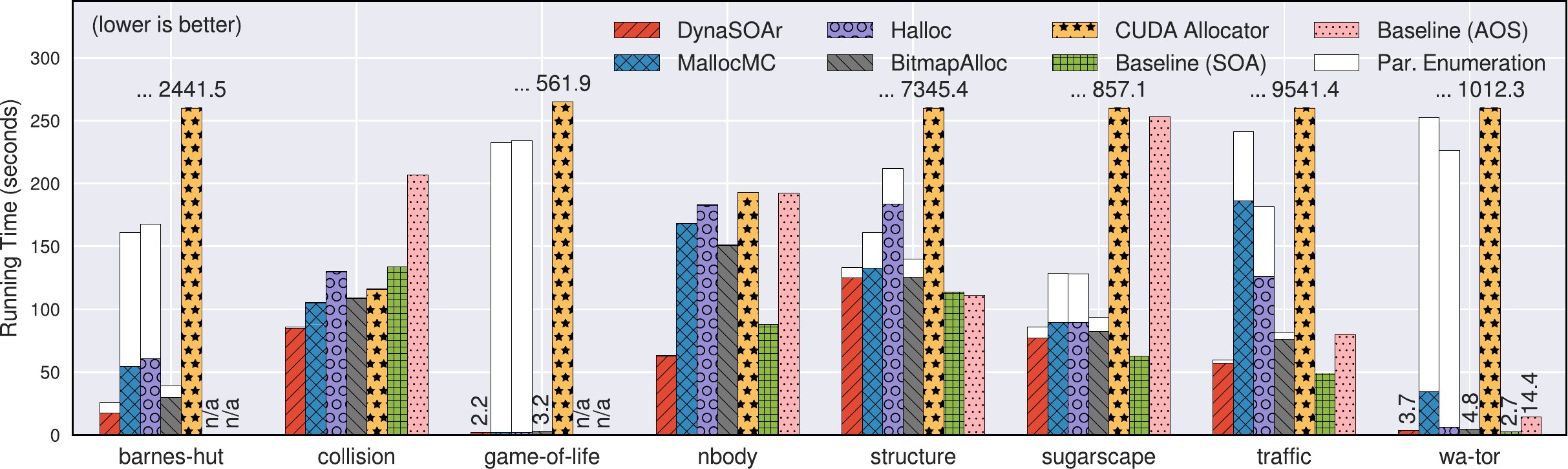}
  \caption{Running time of SMMO applications with different allocators.}
  \label{fig:bench_runntime}
\end{figure*}

\begin{figure*}
  \centering
  \includegraphics[width=0.85\textwidth]{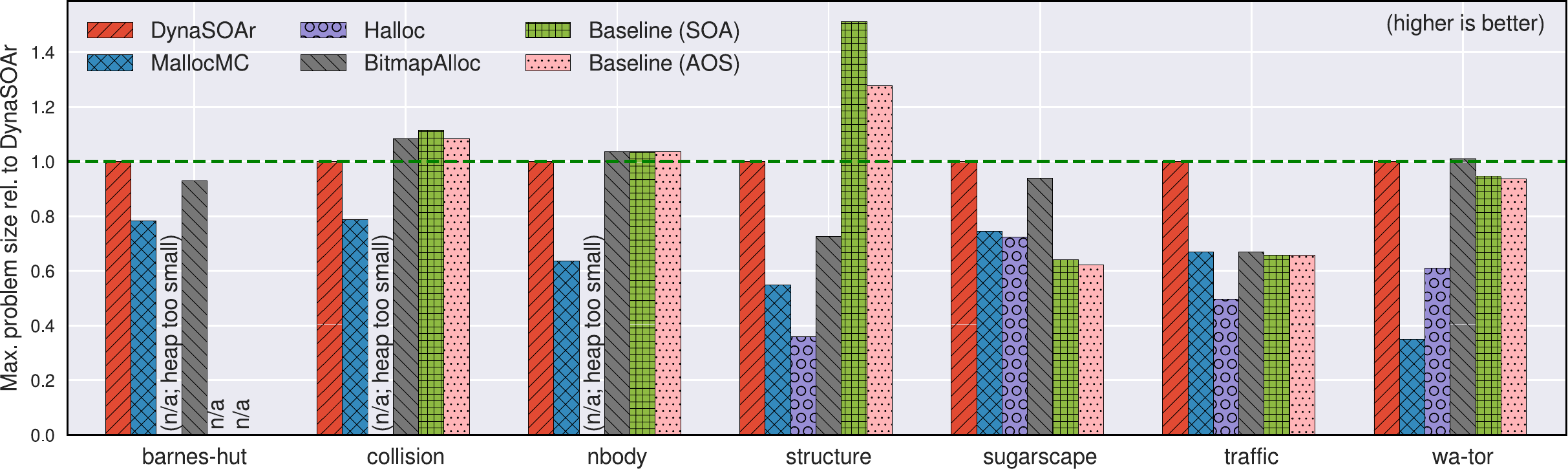}
  \caption{Space efficiency of SMMO applications with different allocators.}
  \label{fig:bench_space_eff}
\end{figure*}
We evaluated the running time (Fig.~\ref{fig:bench_runntime}) and memory usage (Fig.~\ref{fig:bench_space_eff}) of 8 SMMO applications (Fig.~\ref{fig:smmo_apps}) with different allocators on an NVIDIA Titan Xp GPU (12~GB device memory). We compared \textsc{DynaSOAr} with two state-of-the-art GPU allocators (\emph{mallocMC}~\cite{6339604}, \emph{halloc}~\cite{hallocweb}) and with another baseline allocator \textsc{BitmapAlloc} that we developed. \textsc{BitmapAlloc} allocates objects in a large array and uses a bitmap to find empty slots. If possible, we also implemented variants without any dynamic allocation (\emph{Baseline AOS/SOA}).

\paragraph{Running Time}
We break down application running times by the amount of time spent on parallel enumeration (i.e., Fig.~\ref{fig:thread_assignment}) and the remaining running time. Other allocators do not support parallel do-all out of the box, so we had to reimplement it for the benchmarks. With higher engineering efforts we could likely build a more efficient implementation. For a fair comparison of allocators, we should not take parallel enumeration time into account.

All applications except for \textsf{sugarscape} benefit from SOA (compare baselines AOS/SOA). \textsc{DynaSOAr} always exhibits better performance than the other allocators. The CUDA profiler indicated that this is due to better memory coalescing (SOA) in most cases. \textsf{nbody} and \textsf{collision} use little memory, so these applications benefit also from better L1/L2 cache utilization due to \textsc{DynaSOAr}'s compact allocation policy.

\paragraph{Space Efficiency}
To measure how efficient the allocators manage heap memory, we gave every allocator the same amount of memory and experimentally determined the maximum problem size before the application crashes with an out-of-memory error.

MallocMC and Halloc use a hashing approach to find empty memory locations during allocations. This requires little thread synchronization and usually results in fast allocations. However, the performance of every hashing approach decreases with the number of collisions. We believe this is why MallocMC and Halloc can utilize only a much smaller fraction of the heap compared to \textsc{DynaSOAr} and \textsc{BitmapAlloc}.

Many applications (e.g., \textsf{traffic}) cannot be implemented space-efficiently without dynamic allocation because the number of runtime objects of a type cannot be predicted accurately or changes over time.

\section{Conclusion}
\textsc{DynaSOAr} is a CUDA framework for SMMO applications on GPUs. The main insight of our work is that optimizing only for fast (de)allocations is not enough. Optimizing the access of allocated memory can result in much higher speedups, because memory access is the biggest bottleneck of many GPU applications. \textsc{DynaSOAr} achieves this by controlling both data layout (SOA) and data access patterns (parallel do-all), combined with a dense allocation policy that is optimized with hierarchical bitmaps.

\begin{acks}
This work was supported by JSPS KAKENHI Grant Number \grantnum{JSPS}{18J14726} and a Titan Xp hardware donation from NVIDIA Corporation.
\end{acks}

\bibliographystyle{ACM-Reference-Format}
\bibliography{paper}

\end{document}